\documentclass[english]{article}
\usepackage[latin9]{inputenc}
\usepackage{textcomp}
\usepackage{ifsym}
\usepackage{graphicx}
\usepackage{esint}

\makeatletter

\providecommand{\tabularnewline}{\\}

\newcommand{\lyxaddress}[1]{
\par {\raggedright #1
\vspace{1.4em}
\noindent\par}
}

\@ifundefined{showcaptionsetup}{}{%
 \PassOptionsToPackage{caption=false}{subfig}}
\usepackage{subfig}
\makeatother

\usepackage{babel}
\begin{document}

\title{Few-cycle fiber pulse compression and evolution of negative resonant radiation}

\author{J. McLenaghan and F. K\"onig}

\maketitle

\lyxaddress{\begin{center}
\textit{\footnotesize{School of Physics and Astronomy, SUPA, University
of St. Andrews, North Haugh, St. Andrews, KY16 9SS, UK}}
\par\end{center}}
\begin{abstract}
We present numerical simulations and experimental observations of the spectral expansion of fs-pulses compressing in optical fibers. Using the input pulse frequency chirp we are able to scan through the pulse compression spectra and observe in detail the emergence of negative-frequency resonant radiation (NRR), a recently discovered pulse instability coupling to negative frequencies \cite{Rubino:2012ly}. We observe how the compressing pulse is exciting NRR as long as it overlaps spectrally with the resonant frequency. Furthermore, we observe that optimal pulse compression can be achieved at an optimal input chirp and for an optimal fiber length. The results are important for Kerr-effect pulse compressors, to generate novel light sources, as well as for the observation of quantum vacuum radiation.

\end{abstract}

\section{Introduction}

Negative resonant radiation (NRR) is a recently discovered dispersive wave generation process in nonlinear optical  media \cite{Rubino:2012ly}. It tranfers energy from a soliton to a dispersive wave under a phase matched process that couples positive soliton frequencies to
negative frequencies \cite{Rubino:2012fk}. It is a negative-frequency extension to the well known resonant radiation (RR), that couples positive frequencies only  \cite{Akhmediev:1995hc,Wai:86} and is an important part of supercontinuum generation \cite{Alfano:1970,
Dudley:2006nx,
Agrawal:01,
Russell2003}.
RR is also important in various possible
applications including photobiology \cite{Tu:12}, astrocomb generation \cite{Chang:12},
and the production of squeezed states \cite{Tran:2011qf}. 

NRR is particularly interesting as it demonstrates the necessity to include negative frequencies into the desciption of a field in nonlinear optics. 
Whilst negative frequencies are present in all electromagnetic fields
they are typically ignored as they contain the same information as
their corresponding positive frequencies. This can be seen by expressing
a field $A$ by Fourier transform:
\begin{equation}
A\left(z,t\right)=\frac{1}{2\pi}\intop_{-\infty}^{\infty}\widetilde{A}\left(z,\omega\right)e^{-i\omega t}d\omega
\end{equation}
Here $\widetilde{A}\left(z,\omega\right)$ has been integrated over
both positive and negative frequencies. For a real field $\widetilde{A}\left(z,-\omega\right)=\widetilde{A}^{*}\left(z,\omega\right)$, i.e. the negative frequency part is determined by the positive frequency part.
However, the interaction of postive and negative frequency components of a field can lead to qualitatively new phenomena such as NRR. 

The mixing of positive and negative frequencies is a familiar concept in quantum field theory \cite{Birrell1984}. For example, it is behind the
generation of Hawking radiation \cite{Hawking1974,Hawking1975}. Laboratory analogues
of Hawking radiation \cite{Unruh1981} have been of great interest in recent decades
and several different implementations have been suggested to study the effect using e.g.
water waves \cite{Weinfurtner:2011uq}, BECs \cite{Barcelo:2003fk},
optical materials \cite{Philbin:2008fr,Belgiorno:10}, or superconducting
circuits \cite{Nation:2012uq}.  Interestingly, one analogue system also uses solitons in fibers. In this system the
analogue Hawking effect is produced by scattering
of the quantum vacuum at the optical event horizon\cite{Philbin:2008fr}.
NRR can also be described as the scattering of light
at the horizon, however, here bright light from the pulse is scattered \cite{Choudhary:12}. Although more light needs to be shed on the relation between these two effects, it is interesting to notice that they are occuring under similar experimental conditions:
NRR is observed for extreme nonlinear pulse compression \cite{Rubino:2012fk,Akhmediev:1995hc} whilst the analogue
Hawking radiation is expected to occur for extremely steep pulse fronts \cite{Philbin:2008fr}.

In order to measure how the NRR radiation is emerging from a compressing pulse, experimental control of the pulse compression is required. Various tools have been used to control pulse evolution and compression in a fiber inlcuding the tuning of the pulse wavelength, pulse length,
pulse power, and changes in the fiber core size and in the fiber dispersion \cite{Chang:10,Roy:2009kl,Tu:09,Tu:09b}. 
A variation of the pulse power effectively changes where in the fiber the pulse compresses, however, this method has a limited range determined by the onset of NRR and the maximum power available. In addition, the degree of spectral
expansion and generation efficiency of the NRR are significantly affected in a nontrivial way.
Having said that, the frequency chirp of the input pulse can lead to a delay of nonlinear pulse evolution. The dependence on chirp of various aspects of pulse evolution have been considered previously including the effect on supercontinuum generation \cite{Cheng2011,Fu2004,Zhang2007,Zhu2004} and spectral broadening \cite{Tianprateep2005, Tianprateep2004}. In the latter two papers the influence of chirp on spectral broadening was investigated numerically and experimentally for ($50$fs) pulses. The results show an increased spectral broadening at the end of a ($10$cm) piece of fiber when the input pulse chirp is positive and balanced by the total negative dispersion over the fiber length. This occurs because the input chirp must be compensated, delaying pulse compression. 
In this paper we present to the best of our knowledge the first detailed investigation of chirp looking beyond the idea of delayed pulse compression. We look at a regime of shorter pulses ($<20$fs) and shorter fibers ($<0.1$m) and find interesting effects such as the increased degree of pulse compression for a small positive chirp compared to zero chirp. As in the papers mentioned above we see delayed pulse compression and use this as a convenient tool to observe the evolution of both the pulse and the NRR under near constant spectral expansion.  

In this paper, firstly, we present simulations of the pulse evolution and compression. We investigate how the pulse chirp affects the pulse compression. Secondly, we present measurements of the pulse spectra and NRR spectra for variable pulse chirps. We give a detailed discussion of the observations and implications from our simulations. As a result, we observe that the pulse compression has a weak dependence on chirp, however, the chirp is able to delay or advance the pulse compression in the fiber in accordance with the expectation from dispersion cancellation. NRR is generated at the point in the fiber where the pulse starts to compress and generation ceases after compression is over and the pulse expands again. The amount of NRR detected at the end of the fiber is also affected by waveguide loss. Our observations are in qualitative agreement with the analytical description of NRR evolution in fibers  \cite{Akhmediev:1995hc, Rubino:2012fk}.

\section{Resonant radiation}
NRR and RR form as energy is transferred from a pulse, similar to a soliton, propagating
in the anomalous dispersion region of an optical fiber to light in
the normal dispersion region. Light propagation in the fiber is governed
by the generalized non-linear Schr\"odinger equation (GNLSE)\cite{Agrawal:01}. Neglecting absorption and higher order terms, except dispersion, the propagation equation
becomes:
\begin{equation}
\frac{\partial A}{\partial z}-iD\left(i\frac{\partial}{\partial T}\right)A=i\gamma\left|A\right|^{2}A,
\end{equation}
where $A(z, T)$ is the pulse envelope, $\gamma$ is the non-linear parameter of the fiber, and $F.T.[D\left(i\frac{\partial}{\partial T}\right)]=\widetilde{D}(\omega)=\sum_{n\geq2}\frac{\beta_{n}}{n!}\omega^{n}$ represents the second and
higher order-group velocity dispersion. $T=t-z/v_{g}$ is the retarded time in a reference frame moving with the group velocity
$v_{g}$ of the pulse. 

\begin{figure}
\begin{centering}
\begin{tabular}{cc}
\includegraphics[scale=0.35]{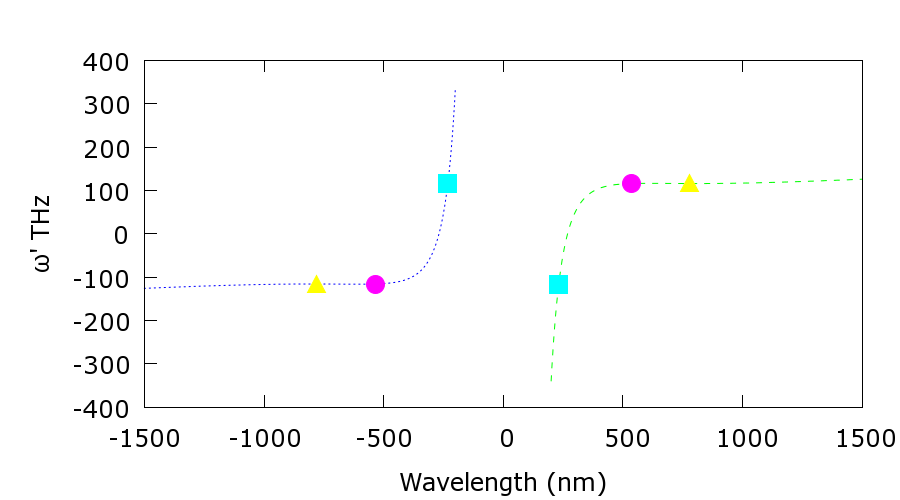} 
\end{tabular}
\par\end{centering}
\caption{Co-moving frame dispersion relation for positive (dashed line) and negative (dotted line) laboratory wavelengths. The soliton (\textifsymbol[ifgeo]{49}),
RR (\textbigcircle{}) and NRR (\textifsymbol[ifgeo]{96}) are indicated. Fiber NL-1.5-590, NKT Photonics Inc. \label{fig:dispZoomOut}}
\end{figure}

\begin{figure}
\begin{centering}
\begin{tabular}{cc}
\subfloat[Positive $\omega'$ \label{fig:neg PM curve}]
{\begin{centering}
\includegraphics[scale=0.175]{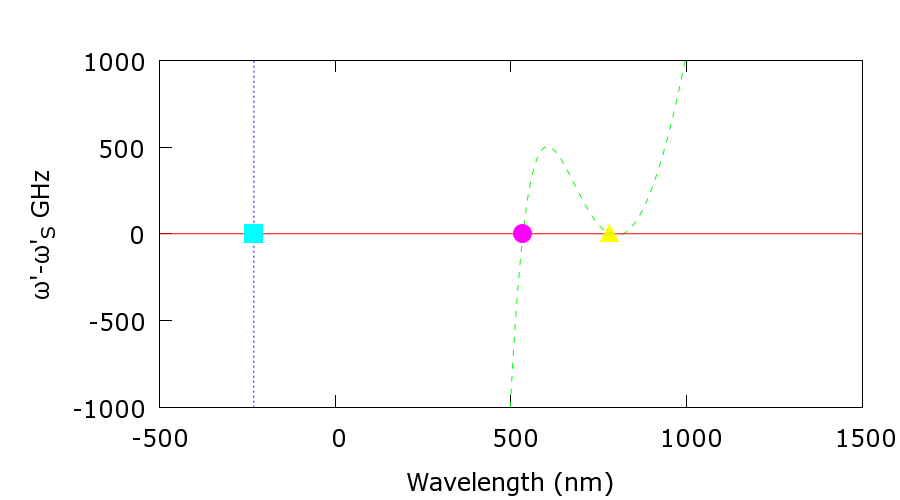}
\end{centering}
}
  &
\subfloat[Negative $\omega'$ \label{fig:CC}]
{\begin{centering}
\includegraphics[scale=0.175]{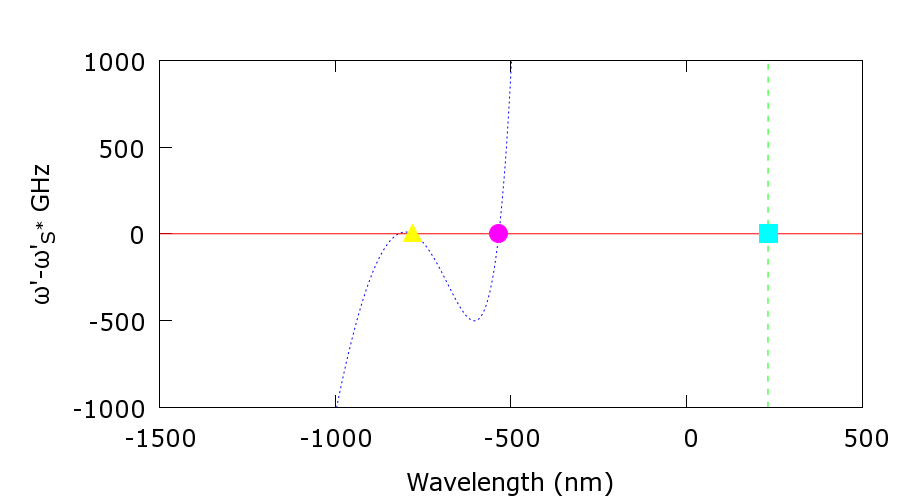}
\end{centering}
}
\tabularnewline
\end{tabular} \par \end{centering}
\caption{Co-moving frame dispersion relation for
positive (dashed line) and negative (dotted line) laboratory wavelengths as in Fig. \ref{fig:dispZoomOut}. Zoomed view around (a)  $\omega'_s$ and (b)  $-\omega'_s$. The soliton (\textifsymbol[ifgeo]{49}),
RR (\textbigcircle{}) and NRR (\textifsymbol[ifgeo]{96}) are indicated.
 \label{fig:Co-moving-frame-frequency+neg}}
\end{figure}

An important parameter characterising soliton amplitudes is the soliton order $N$, given by $N^{2}=\gamma P_{0}T_{0}^{2}/\left|\beta_{2}\right|$ where  $P_{0}$ and $T_{0}$ are the soliton peak power and length, respectively. This parameter can also be used for soliton-like pulses in our experiments. For higher ($N>1$) order solitons the instabilities caused by higher order effects lead to fission into $N$ fundamental ($N=1$) solitons \cite{Herrmann:2002fj,Husakou:2001rt}.  
The higher order dispersion is perturbing each of the produced solitons, which consequently will radiate dispersive RR and NRR. 
The radiation generated at different times as the pulse propagates
along the fiber interferes constructively if the following momentum conserving phase matching condition is
satisfied\cite{Cristiani:04}:
\begin{equation}
\sum_{n\geq2}\frac{\beta_{n}}{n!}\left(\omega-\omega_{\mathrm{P}}\right)^{n}=\frac{\left(2N-1\right)^{2}\gamma P_{0}}{2N^{2}}\label{eq:PM0},
\end{equation}
where $\omega$ and $\omega_{\mathrm{P}}$ are the phase matched radiation (RR or NRR) and pulse 
frequencies, respectively. 
In order to graphically find the phase matched frequencies we express Eq. \ref{eq:PM0} as a frequency conservation in the co-moving frame \cite{Rubino:2012ly,Philbin:2008fr}: 
\begin{equation}
\omega'_{\mathrm{P}}=\omega'_{\mathrm{NRR}}\label{eq:PM}
\end{equation}
where $\omega'$ is the (Galilean) co-moving frame
frequency of the pulse and NRR, respectively, given by $\omega'=\omega-v_{g}(\beta(\omega)+\beta_{nl})$. $\beta$ is related to the refractive index by $\beta=\frac{n\left(\omega\right)\omega}{c}$ and $\beta_{nl}$ is the nonlinear contribution to the propagation constant due to the soliton pulse. 
Figure \ref{fig:dispZoomOut} shows the dispersion relation of our fiber in the co-moving frame ($\beta_{nl}=0$). Figure \ref{fig:Co-moving-frame-frequency+neg}a shows a closer view of the top part of the dispersion relation around the soliton co-moving frame frequency $\omega'_s$. Figure \ref{fig:Co-moving-frame-frequency+neg}b shows the lower part of the dispersion relation plotted around  $-\omega'_s$.  There are two branches to the relation seen in each of the three plots, one with positive and one with negative laboratory frequency (wavelength). Marked with a symbol in Fig. \ref{fig:neg PM curve}  are the soliton, the resonant radiation wavelength (RR), and a third solution, the negative resonant radiation (NRR) on the respective other branch. In Fig. \ref{fig:CC} the complex conjugate fields are displayed. This shows that the soliton can excite two other waves, each of which is a real-numbered field, one with a positive and one with a negative frequency (wavelength). 


\section{Numerical simulations}
In order to understand the pulse evolution, i.e. the pulse compression and spectral expansion, under the influence of pulse chirp we numerically simulated the pulse amplitude propagation using a split step Fourier simulation tool \cite{Paschotta}.

We used a $12\:\mathrm{fs}$ (FWHM) hyperbolic secant pulse centered at $800\:\mathrm{nm}$ wavelength with the amplitude of an $\mathrm{N}=2.25$
soliton similar to the experiment. 
The frequency chirp of the initial pulse was varied and the evolutions
of the pulse spectrum and the peak power were recorded. Because we chose a highly nonlinear fiber, the pulse compresses over a few $\mathrm{mm}$. Hence effects taking place
over longer propagation distances such as stimulated Raman scattering are not
considered in detail. We simulated the propagation of pulses through a fiber
with the dispersion profile shown in Figs. \ref{fig:dispZoomOut} and \ref{fig:Co-moving-frame-frequency+neg}.

Pulse compression in an optical fiber occurs due to the interplay
between self-phase modulation (SPM) and group velocity dispersion
(GVD). SPM broadens the pulse spectrum and generates a positive chirp, i.e. long wavelengths lead and short wavelengths trail. Anomalous GVD slows the long wavelengths and speeds up short wavelengths, resulting in pulse compression \footnote{Compression only occurs across the central region
of the pulse where both the SPM and GVD induced chirps are approximately
linear.}. Adding, via pulse dispersion, a frequency chirp to the input pulse has two key effects on
the pulse compression. Firstly, the distance at which the pulse compresses to its shortest length is modified. If the initial chirp
is positive, then the anomalous fiber GVD acts to reduce the chirp and increase the pulse peak power before the nonlinear dynamics fully compresses the pulse. Hence positive input chirp delays compression further along the fiber. Secondly, the amount by which the pulse
compresses can be either enhanced or reduced.

\begin{figure}
\begin{centering}
\begin{tabular}{ccc}
\includegraphics[scale=0.2]{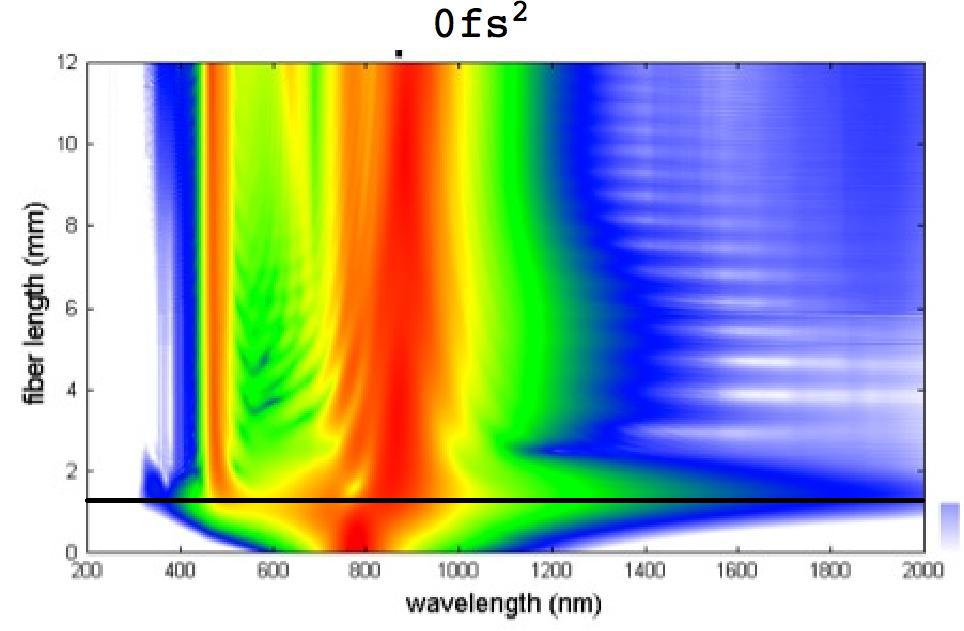} & \includegraphics[scale=0.2]{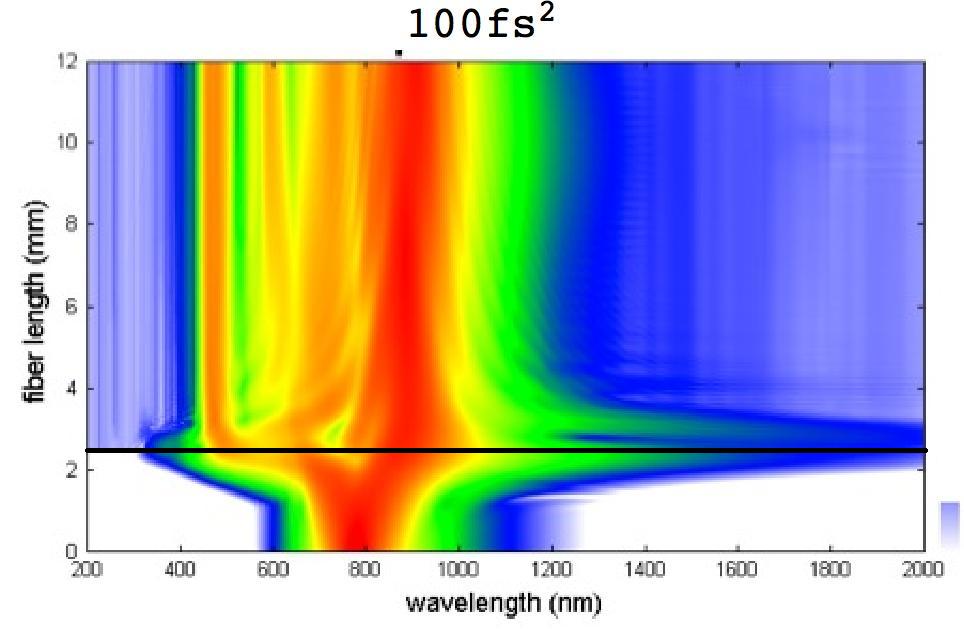} \tabularnewline
 \includegraphics[scale=0.2]{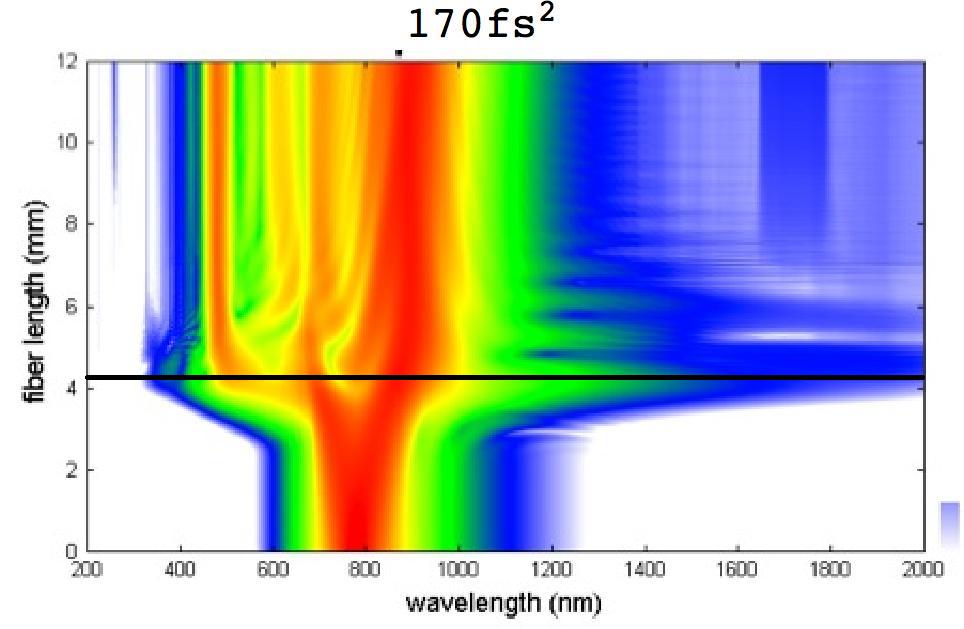} & \includegraphics[scale=0.2]{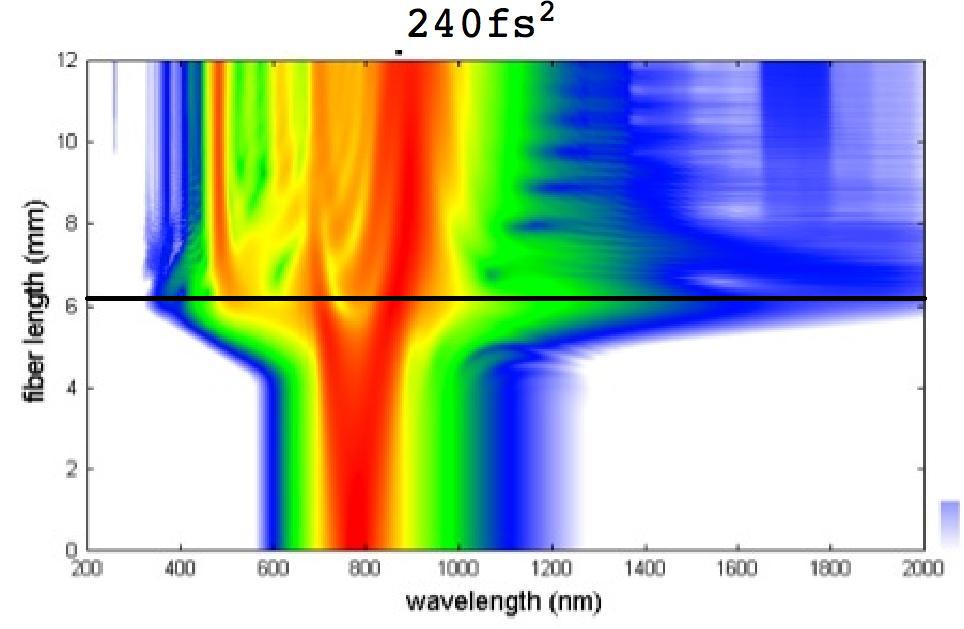}\tabularnewline
\includegraphics[scale=0.2]{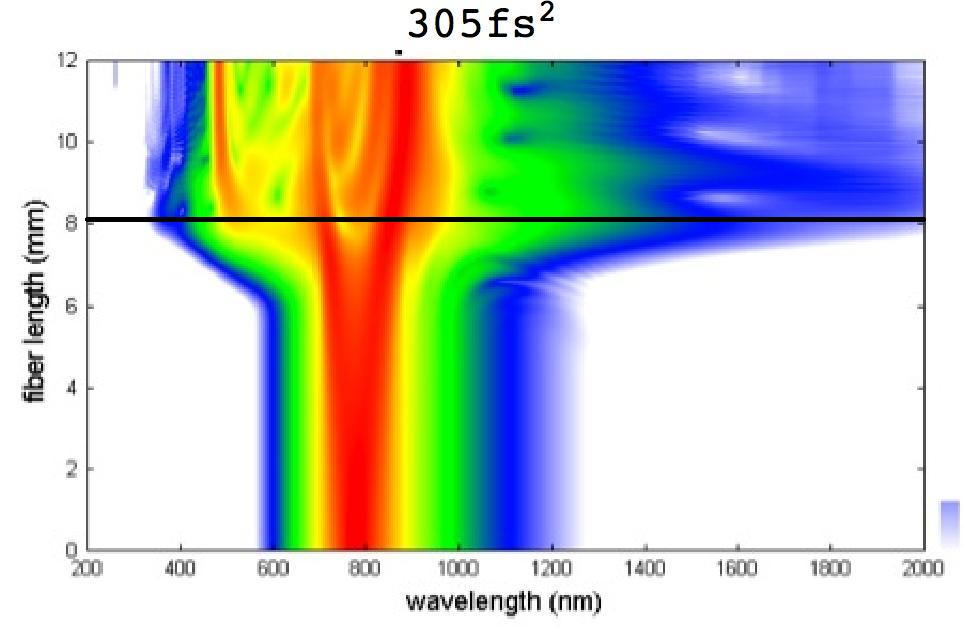} & \includegraphics[scale=0.2]{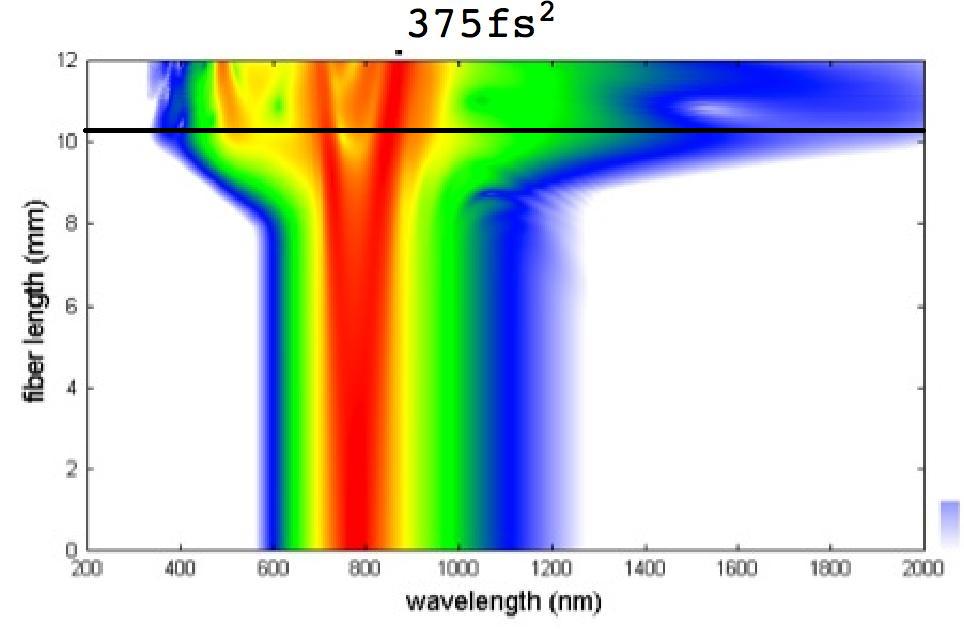}\tabularnewline
\end{tabular}
\par\end{centering}
\caption{Pulse spectral evolution during propagation
for different input pulse chirps. The horizontal line indicates the maximum spectral broadening. 
\label{fig:ContourSim}}
\end{figure}

\begin{figure}
\begin{centering}
\includegraphics[scale=0.4]{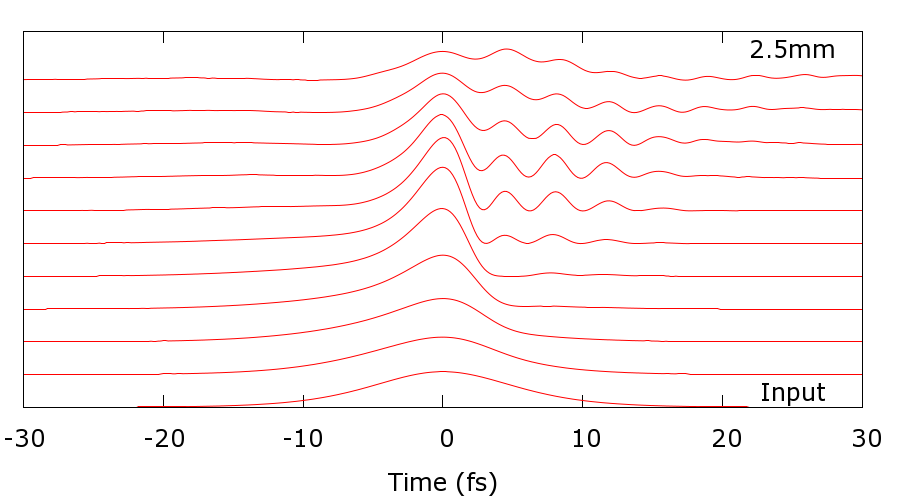}
\par\end{centering}
\caption{Evolution of the pulse temporal envelope as the pulse propagates from
$0\:\mathrm{mm}$ to $2.5\:\mathrm{mm}$ along the fiber. The pulse
has zero initial chirp.\label{fig:temporal evolution}}
\end{figure}

We started our simulations to verify this qualitative behaviour. In Fig. \ref{fig:ContourSim} a series of
spectral evolutions is displayed for different input chirps (in units of $\mathrm{fs^{2}}$). For comparison, the fibers we use in experiments have dispersions between $20 \mathrm{\: fs^{2}}/mm$ and $40 \mathrm{\: fs^{2}}/mm$. If no chirp is applied, the initial pulse spectrum
centered in the near infrared can be seen to expand over three octaves
within the first $2\mathrm{\: mm}$ of propagation. The spectrum then
contracts again at $2.5\mathrm{\: mm}$ as the pulse continues to propagate. This
corresponds to the temporal evolution of the pulse given in Fig. \ref{fig:temporal evolution}.
If a positive input chirp is applied, Fig. \ref{fig:ContourSim} shows that the distance at
which the pulse compresses and the spectrum expands increases accordingly.
In addition, the extent of spectral broadening reduces. 

\begin{figure}
\begin{centering}
\begin{tabular}{cc}
\includegraphics[scale=0.375]{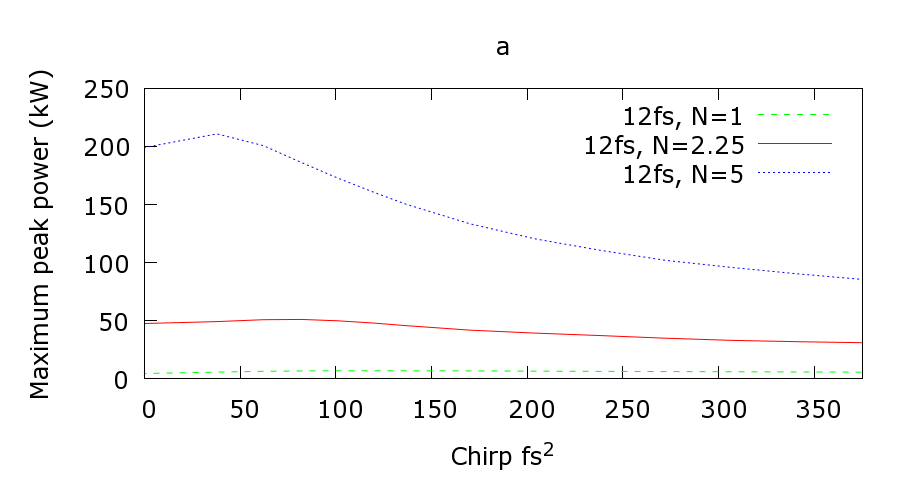} \tabularnewline
 \includegraphics[scale=0.375]{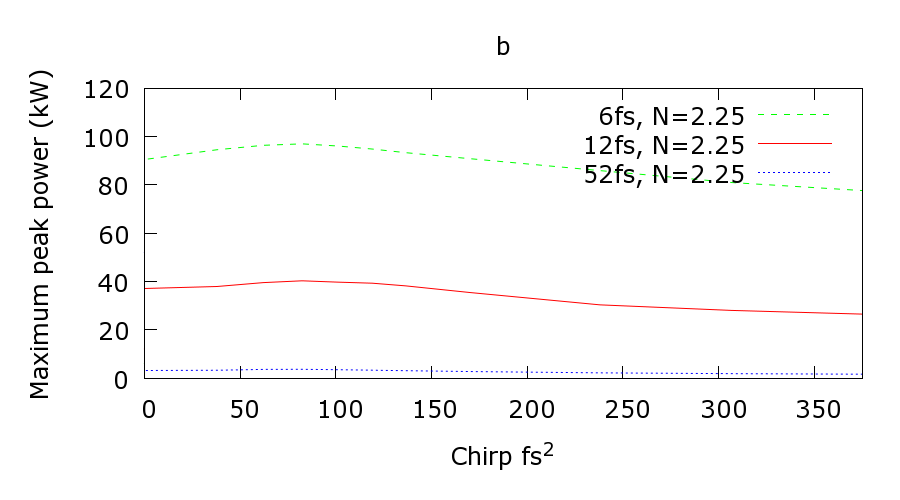}\tabularnewline
\end{tabular}
\par\end{centering}
\caption{Maximum peak power during pulse compression as a function of input chirp. Results are for different input pulse soliton orders (a: $N=5$ dotted; $N=2.25$ solid; $N=1$ dashed) and pulse lengths (b: $6\:\mathrm{fs}$ dashed; $12\:\mathrm{fs}$  solid; $52\:\mathrm{fs}$  dotted). \label{fig:Maximum-peak-power} }
\end{figure}

\begin{figure}
\begin{centering}
\includegraphics[scale=0.375]{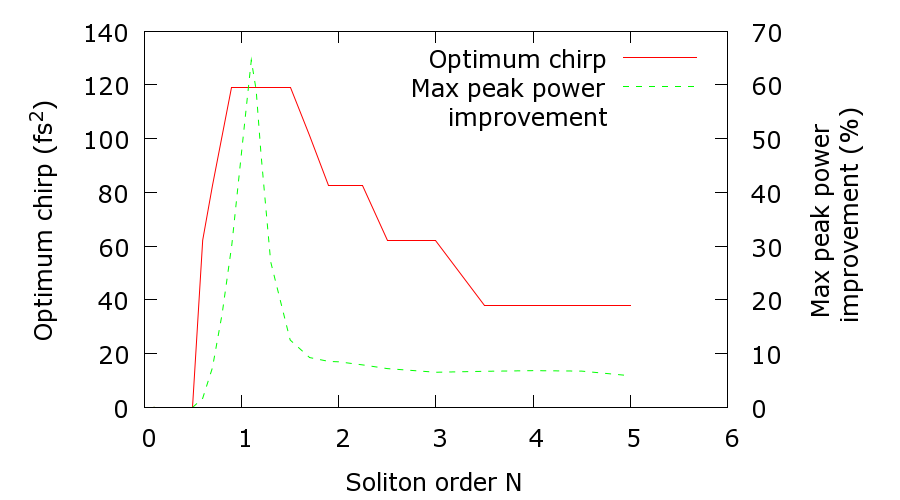}
\par\end{centering}
\caption{Optimum chirp for pulse compression as a function of soliton order
$\mathrm{N}$; maximum peak power enhancement using the optimum chirp
compared to zero chirp.\label{fig:Maximum-peak-powermanyN}}
\end{figure}

In the time domain, as shown in Fig. \ref{fig:temporal evolution}, the peak power first increases to a maximum at the maximal spectral expansion and then reduces again as the pulse broadens in time. The dependence of this maximum peak power on input
chirp is displayed in Fig. \ref{fig:Maximum-peak-power} for a set of
input pulse parameters. The solid line in each of the sub-figures
represents a $12\:\mathrm{fs}$, $\mathrm{N}=2.25$
input pulse. Clearly, the peak power increases with amplitude $N$ and reduces with pulse duration. Also, the maximum peak power decreases eventually for large chirp for all amplitudes and pulse durations shown. For all traces in Fig. \ref{fig:Maximum-peak-power} there is a finite optimal input chirp that leads to maximum pulse compression. The enhancement seems to be best for the $12\:\mathrm{fs}$ pulse duration. Fig.  \ref{fig:Maximum-peak-powermanyN} shows how this non-zero chirp depends on pulse amplitude. For all amplitudes, there is a positive chirp leading to an improvement of the pulse compression, which is particularly pronounced for smaller pulse amplitudes. For the  $\mathrm{N}=2.25$ pulse we also increased the pulse length and found the optimum chirp to slightly decrease from $80\:\mathrm{fs^{2}}$ to a steady
value of around $60\:\mathrm{fs^{2}}$. The improvement in the maximum peak power rose to a maximum of $18\:\%$ for a $50\:\mathrm{fs}$ pulse (results not shown).  
For our fiber a chirp of $375\:\mathrm{fs^{2}}$ would change the distance of compression from close to zero to the end of our longest fibers of $14\:\mathrm{mm}$ length. According to Fig. \ref{fig:Maximum-peak-power}, over the same range the maximum peak power increases first by $9\%$ and then falls to $36\%$ below the unchiped compression. This means that as the compression is delayed along the fiber, there is only a small variation in compression efficiency. Effectively, the pulse compression can be observed in its different stages at the fiber end by small variations of the pulse chirp around an offset chirp compensating the fiber dispersion. If instead the pulse power were used to move the compression point from the start to the end of the fiber, the variation in power required be considerably larger for similar pulse parameters. For example, for an input chirp of $240\:\mathrm{fs^{2}}$ and a $12\mathrm{\: fs}$ pulse, the input power would have to vary by a factor of $25$. 

In addition to investigating the effects of  input chirp, pulse power, and pulse
length on pulse compression we turn on and off higher order effects such as self steepening and the Raman
effect. We find that the self-steepening has a negligible
effect on the compression distance and slightly reduces both, the
spectral expansion and the maximum peak power. The Raman effect increases slightly the propagation
distance before the compression point but has a negligible effect on the spectral expansion or the maximum peak power. 

Summarizing, the simulations show that, for a wide range of input pulse parameters, chirp can be used to move
the pulse compression along the fiber.  The accompanying changes in the degree of pulse compression must be taken into account but are small enough to allow for a qualitative 
investigation of the pulse and NRR evolution. The origin of these changes is not fully understood and is to be investigated elsewhere. Furthermore, a small positive input chirp leads to maximal pulse compression. 

\section{Experiment}
 
We investigate two principal aspects of the pulse and NRR evolution. The first is the pulse spectral expansion. In order to
excite the NRR in the UV (Fig. \ref{fig:Co-moving-frame-frequency+neg}), the driving input pulse in the IR must broaden its spectrum extensively to reach the NRR wavelength. The spectral broadening is a transient effect during the pulse compression. Therefore, in order to measure the extent of the broadening into the UV we use the input chirp to move each stage of the pulse compression, respectively, to the fiber end. The second aspect is the evolution of the NRR itself, which is generated when the pulse compresses. By moving the compression point again we can change the distance propagated by the NRR between its first generation
and the fiber end and observe its evolution. In addition we see how the variation in compression efficiency with chirp affects the generation of NRR. 
\begin{figure}
\begin{centering}
\includegraphics[scale=0.375]{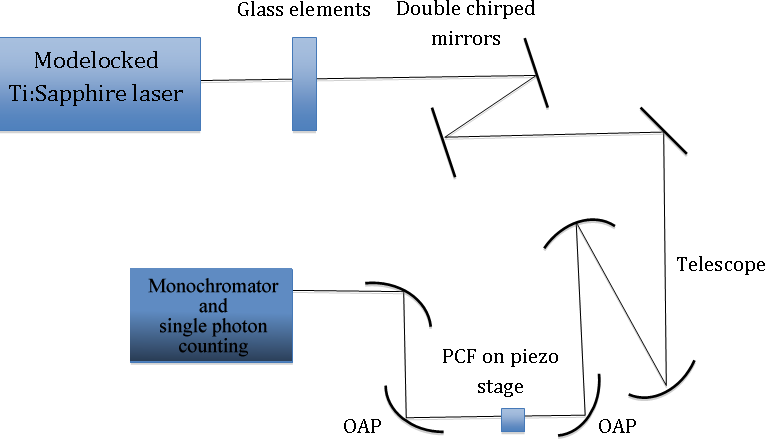}
\par\end{centering}
\caption{Experimental setup: BK7 glass plates and wedges were used to control
the input pulse chirp. \label{fig:Experimental-setup}}
\end{figure}

Figure \ref{fig:Experimental-setup} shows  the experimental setup. We use $800\mathrm{\: nm}$-pulses from a mode-locked Ti:Sapphire
laser (FemtoSource Rainbow). A short pass filter at $695\:\mathrm{nm}$ cleans up
the spectrum, resulting in a pulse length (FWHM) of $12\:\mathrm{fs}$
for the unchirped pulse and a bandwidth of over $200\:\mathrm{nm}$. 
The pulses are coupled into a photonic crystal fiber (PCF) (Blaze Photonics NL-1.5-590) with a zero-dispersion
wavelength at about $685\mathrm{\: nm}$. The laser output is linearly polarized and all PCFs are rotated so
as to excite only one polarization axis. The laser wavelength is situated
in the anomalous dispersion region of the PCF which allows for the
formation of solitons. A dispersion optimised coupling into the PCF is achieved using
off-axis parabolic mirrors. UV spectra are recorded using a monochromator with a photomultiplier tube. 
Dispersion is managed up to second order by glass elements in the beam and a pair of doubly chirped mirrors (DCMs).

The dispersive pulse chirp $C$ leads to an increase in the pulse length
and a variation in the instantaneous frequency across the
pulse. It is quantified by $C=\sum_{i}\beta_{2 i}z_{i}+\mathrm{GDD}_{\mathrm{DCM}}$, where $\beta_{2 i}$ and $z_{i}$ are the GVD and
beam path distances for each dispersive optical element before the
fiber and $\mathrm{GDD}_{\mathrm{DCM}}$ is the negative group delay dispersion (GDD)
introduced by the DCMs.

We use different section lengths of PCF from $1\:\mathrm{mm}$ to $14\:\mathrm{mm}$.
For the $1\:\mathrm{mm}$ and $2\:\mathrm{mm}$ lengths the laser power is $100\:\mathrm{mW}$, for the other lengths it is $80\:\mathrm{mW}$. The coupling efficiency into the PCF is around $25\:\%$, varying between fiber samples. 

\section{Spectral expansion}
\begin{figure}
\begin{centering}
\begin{tabular}{cc}
\includegraphics[scale=0.225]{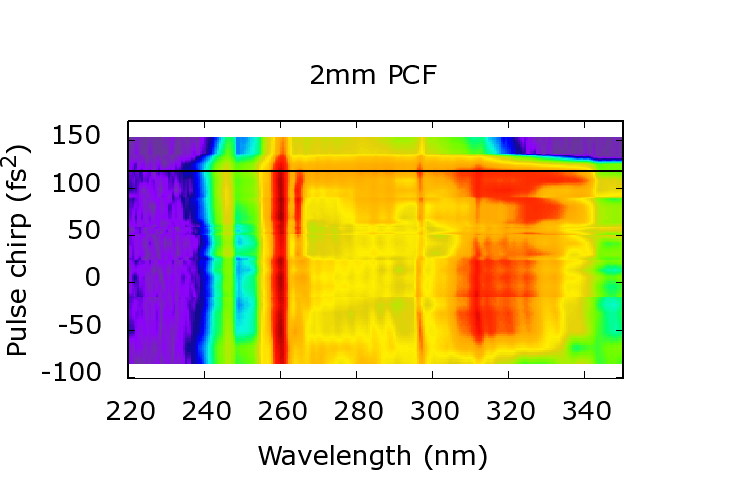} & \includegraphics[scale=0.225]{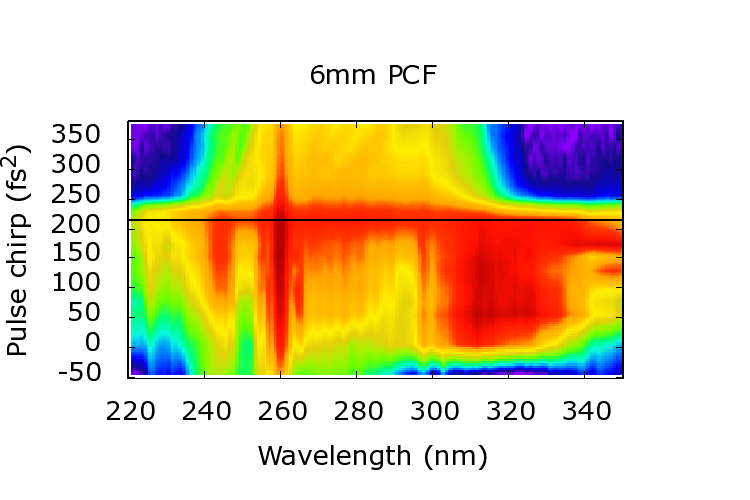}\tabularnewline
\includegraphics[scale=0.225]{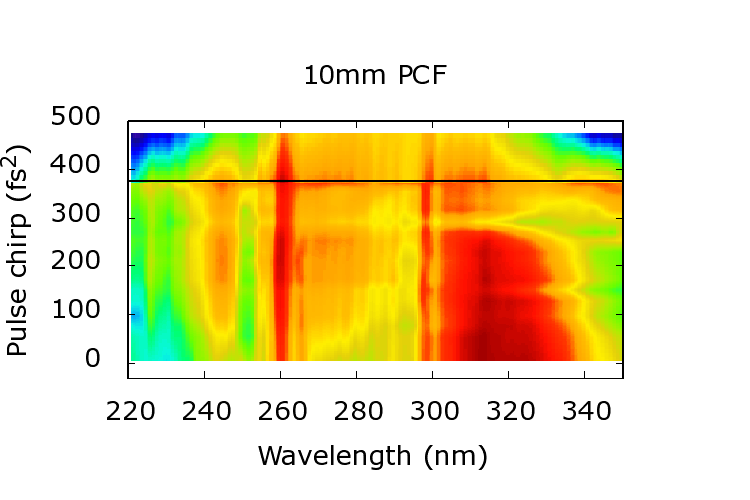} & \includegraphics[scale=0.225]{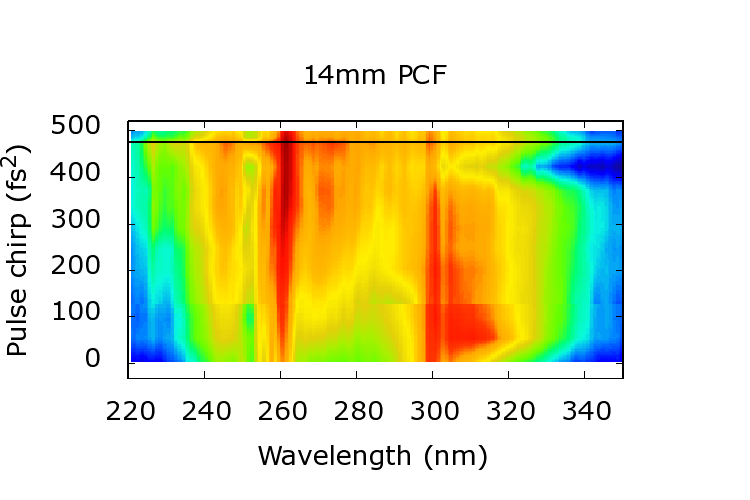}\tabularnewline
\end{tabular}
\par\end{centering}
\caption{Contour plots of spectra as the input pulse chirp
is varied, for different fiber lengths. The horizontal line indicates the largest spectral broadening due to pulse compression. \label{fig:Contour-plots-forUV}}
\end{figure}
The contour plots in Fig.  \ref{fig:Contour-plots-forUV} show the
UV spectra as a function of the input chirp for four
fiber lengths. For each length of fiber there is a particular
input chirp (indicated by a black line) for which the spectrum becomes maximally broad. 
For longer fibers this broadening is observed for higher input chirps. It is a transient effect that disappears at other chirps, demonstrating the transient spectral expansion as the pulse compresses. Interestingly this shows that the pulse broadens its spectrum from the IR to the UV by self-compression.
\begin{table}
\begin{centering}
\begin{tabular}{|c|c|c|c|c|}
\hline 
PCF & Linear & Full&UV Expt.  &NRR Expt. \tabularnewline
 length &  & simulation &  (Fig. \ref{fig:Contour-plots-forUV}) & (Fig. \ref{fig:NRR-plots})\tabularnewline
\hline 
\hline 
$2mm$ & $54\mathrm{}\pm13\:\mathrm{fs^{2}}$ & $96\pm22\:\mathrm{fs^{2}}$ & $117\mathrm{\: fs^{2}}$ & $90\:\mathrm{fs^{2}}$\tabularnewline
\hline 
$6mm$ & $161\mathrm{}\pm13\:\mathrm{fs^{2}}$ & $234\pm30\mathrm{\: fs^{2}}$ & $214\:\mathrm{fs^{2}}$ & $154\:\mathrm{fs^{2}}$\tabularnewline
\hline 
$10mm$ & $269\mathrm{}\pm13\:\mathrm{fs^{2}}$ & $376\pm44\mathrm{\: fs^{2}}$ & $374\:\mathrm{fs^{2}}$ & $374\:\mathrm{fs^{2}}$\tabularnewline
\hline 
$14mm$ & $377\mathrm{}\pm13\:\mathrm{fs^{2}}$ & $518\pm58\:\mathrm{fs^{2}}$ & $474\:\mathrm{fs^{2}}$ & $474\mathrm{\: fs^{2}}$\tabularnewline
\hline 
\end{tabular}
\par\end{centering}
\caption{Comparison of input pulse chirps leading to pulse compression at the end of the fiber as deduced from: linear optics chirp compensation $L\beta_{2}$; numerically simulated pulse compression; observed UV spectral broadening; observed generation of NRR. \label{tab:lengthxGVD}}
\end{table}
Tab. \ref{tab:lengthxGVD} compares input pulse chirps that lead to observable pulse compression at the fiber end, based on different mechanisms: 'Linear' indicates the chirp that would compensate the linear dispersion of the full length of fiber. In order to facilitate comparison with experimental results the errors indicate a range of values for slightly longer or shorter fibers reflecting the experimental uncertainty in the fiber length. The principal nonlinear effect in fibers is SPM, which broadens the spectrum, leading to a faster compression in anomalous dispersion. Thus our numerical simulations of the nonlinear behaviour result in larger input chirps for compression at the fiber end. As previously mentioned the simulations were run using different input powers. Increasing the power enhances the effect of SPM and results in an even larger input chirp for compression at the fiber end. The error values for the simulation results in table \ref{tab:lengthxGVD} indicate how the chirp value varies for a small range of input pulse powers. This allows comparison between simulation and experiment by taking into account the uncertainty in the exact experimental coupling efficiency into each piece of fiber. Next, we compare this chirp to the one used to observe the broadband expansion into the UV. Simulation and experimental results agree well within the errors.

In all of the contour plots of Fig. \ref{fig:Contour-plots-forUV} there are strong signals around $260\:\mathrm{nm}$ and $300-340\:\mathrm{nm}$, independent of input chirp. These are due to third harmonic generation, phase matched between the fundamental and higher order modes\cite{Efimov:03}.  

\section{Pulse compression and generation of NRR}
\begin{figure}
\begin{centering}
\begin{tabular}{cc}
\includegraphics[scale=0.275]{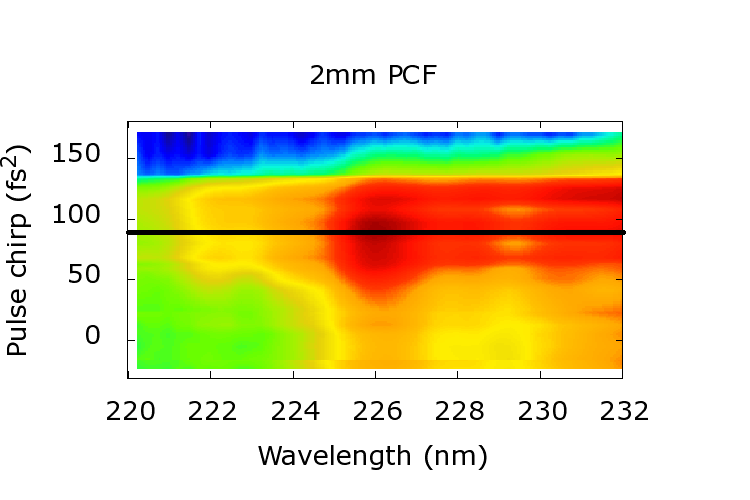} & \includegraphics[scale=0.275]{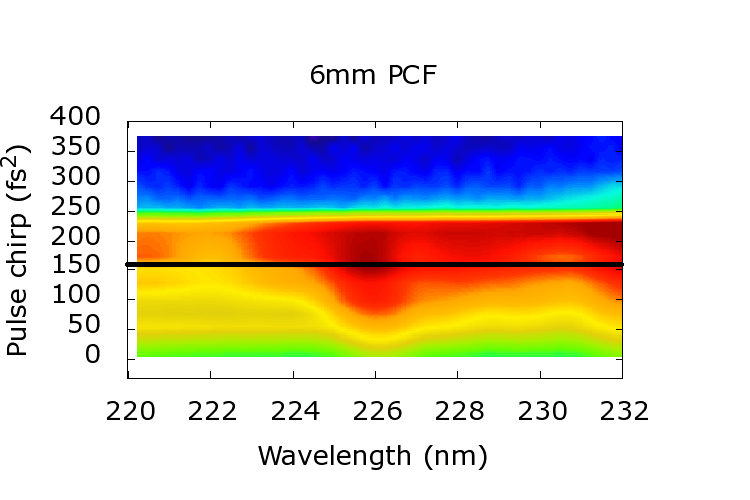}\tabularnewline
\includegraphics[scale=0.275]{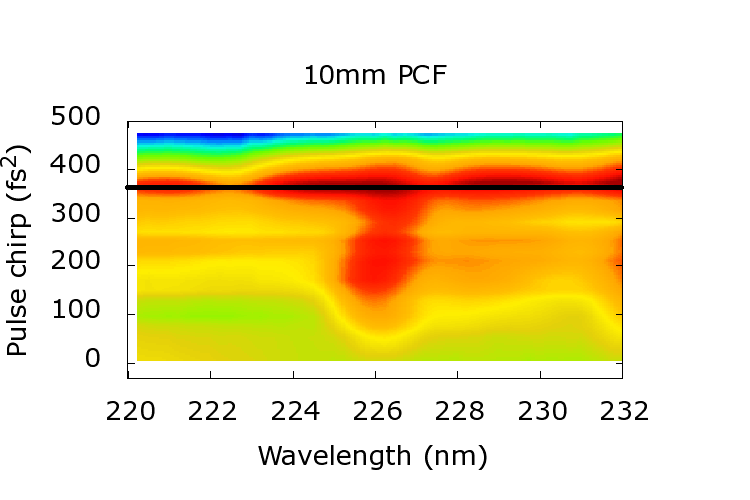} & \includegraphics[scale=0.275]{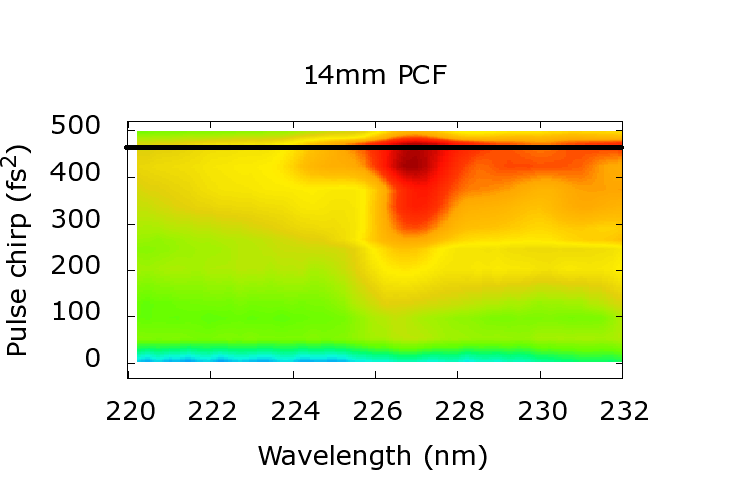}\tabularnewline
\end{tabular}
\par\end{centering}
\caption{Contour plots showing the NRR as the input chirp is varied, the peak
at approximately $226\:\mathrm{nm}$ is the NRR. \label{fig:NRR-plots}}
\end{figure}

In the previous section we have found that a compressing incoming pulse undergoes spectral expansion which reaches far into the UV. In particular, it is reaching the wavelength of negative resonant radiation (NRR), thus allowing an excitation of NRR. Figure \ref{fig:NRR-plots} shows contour
plots of UV spectra around the NRR wavelength of $226\:\mathrm{nm}$, as the input pulse chirp is varied,
for four fiber lengths. 

\begin{figure}
\begin{centering}
\includegraphics[scale=0.4]{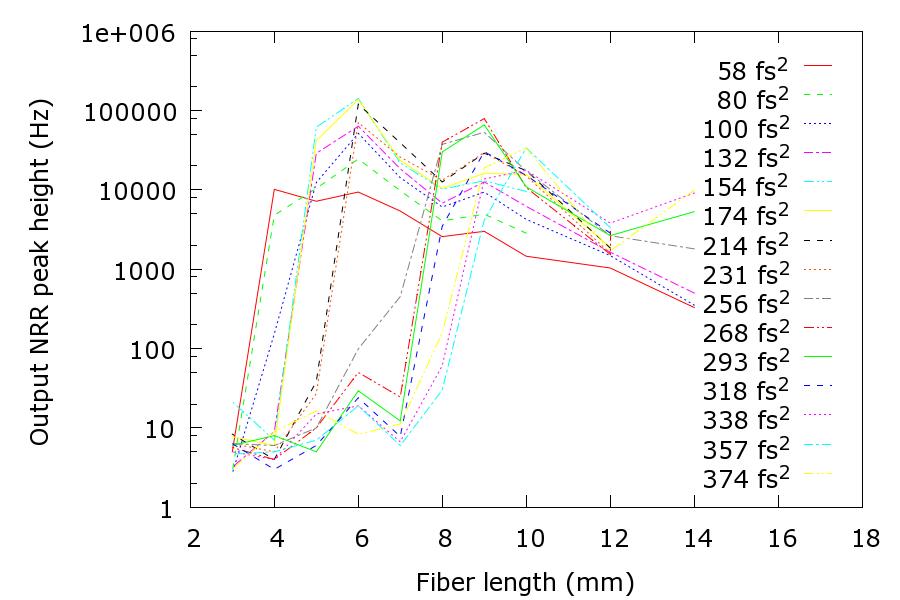}
\par\end{centering}
\caption{NRR production along the fiber under various input chirps. NRR is most efficiently produced around $6\,\mathrm{mm}$ and decays rapidly along the fiber.\label{fig:NRR height}}
\end{figure}

For all fiber lengths the NRR peak is small for negative and
small positive chirps, where the pulse compresses efficiently at the start of fiber propagation. Then it grows to a maximum for a particular chirp, before decreasing. Unsurprisingly, no NRR is generated beyond a chirp leading to pulse compression at the very end of the fiber. The chirp at which the NRR peak reaches its maximum is indicated by a black line in Fig. \ref{fig:NRR-plots} and its numerical value is listed in Tab. \ref{tab:lengthxGVD}. For the two longer fibers, maximum NRR occurs approximately for compression at the end of the fiber instead of at the much smaller chirp of maximum compression (see Fig. \ref{fig:Maximum-peak-powermanyN}). In the case of the
two shorter fibers the maximum NRR is observed for a chirp for compression just before the end of the fiber.
This can be explained by Fig. \ref{fig:NRR height}, which displays the observed NRR signal strength at different fiber lengths for a range of input pulse chirps. Depending on chirp, the pulses compress in the fiber at various distances. It is only during the compression that significant amounts of NRR are generated, after which the UV radiation in the fiber quickly decays. From these data we estimate a UV loss coefficient at  $226\:\mathrm{nm}$ of $1.5-2.5\:\mathrm{dB/mm}$. It is also visible in the figure that NRR production further along the fiber is less efficient, with the optimal fiber length close to $6\:\mathrm{mm}$. This again is a result of the less efficient spectral expansion, cf. Fig.  \ref{fig:ContourSim}.\\

In conclusion we see that it is possible to use the input pulse
chirp as a tool to investigate pulse propagation along a fiber. In particular, we observe pulse compression and the evolution of NRR. By delaying the pulse compression to the end of the fiber
we observe spectral broadening into the UV which is required to excite the NRR. Significant NRR is observed only if the pulse compresses near the fiber end because of fiber loss in the UV. In both the simulations
and the experiments the degree of pulse compression decreases as the input chirp becomes large and positive. Thus we measure the strongest NRR output for a short fiber, at a chirp that simultaneously optimizes pulse compression and delays the compression towards the fiber end. 

The simulations indicate that the deterioration in the pulse compression
with increasing chirp is a persistent feature for a variety of input
pulse powers and pulse lengths. Therefore choosing an appropriate
combination of short fiber and input chirp is crucial in any experiment
to generate light in the UV. This is important for a variety of fs-pulsed experiments such as the creation of supercontinua, the generation of resonant radiation at short wavelength, or the observation of quantum vacuum radiation from artificial optical event horizons \cite{Price2003, Tu:09b,Robertson2011}. 

Other radiation processes based on ultrashort pulse compression can be investigated using the same method, such as the generation of resonant radiation (RR). \\

We would like to acknowledge useful discussions with F. Biancalana, R. Paschotta, and S. Kehr.

\bibliographystyle{plain}
\bibliography{GDD_paper_bibliography}

\begin{thebibliography}{10}

\bibitem{Agrawal:01}
Govind~P. Agrawal.
\newblock {\em Nonlinear Fiber Optics}.
\newblock Optics and Photonics. Academic Press, 3rd edition, 2001.

\bibitem{Akhmediev:1995hc}
Karlsson~Magnus Akhmediev, Nail.
\newblock Cherenkov radiation emitted by solitons in optical fibers.
\newblock {\em Physical Review A}, 51(3):2602--2607, 03 1995.

\bibitem{Alfano:1970}
R.~R. Alfano and S.~L. Shapiro.
\newblock Observation of self-phase modulation and small-scale filaments in
  crystals and glasses.
\newblock {\em Phys Rev Lett}, 24:592--594, 1970.

\bibitem{Barcelo:2003fk}
Carlos Barcel{\'o}, Stefano Liberati, and Matt Visser.
\newblock Towards the observation of hawking radiation in bose--einstein
  condensates.
\newblock {\em International Journal of Modern Physics A}, 18(21):3735--3745,
  2003.

\bibitem{Chang:10}
Guoqing Chang, Li-Jin Chen, and Franz~X. K\"{a}rtner.
\newblock Highly efficient cherenkov radiation in photonic crystal fibers for
  broadband visible wavelength generation.
\newblock {\em Opt. Lett.}, 35(14):2361--2363, Jul 2010.

\bibitem{Chang:12}
Guoqing Chang, Chih-Hao Li, A.~Glenday, G.~Furesz, N.~Langellier, Li-Jin Chen,
  M.W. Webber, Jinkang Lim, Hung-Wen Chen, D.F. Phillips, A.~Szentgyorgyi, R.L.
  Walsworth, and F.X. Kartner.
\newblock Spectrally flat, broadband visible-wavelength astro-comb.
\newblock In {\em Lasers and Electro-Optics (CLEO), 2012 Conference on}, pages
  1--2, 2012.

\bibitem{Cheng2011}
Chunfu Cheng, Youqing Wang, and Qinghua Lv.
\newblock Effect of initial frequency chirp on the supercontinuum generation in
  all-normal dispersion photonic crystal fibers.
\newblock In {\em Photonics and Optoelectronics Meetings (POEM) 2011: Optical
  Communication Systems and Networking}, volume 8331, pages 83310O--83310O--8,
  2011.

\bibitem{Choudhary:12}
Amol Choudhary and Friedrich K\"{o}nig.
\newblock Efficient frequency shifting of dispersive waves at solitons.
\newblock {\em Opt. Express}, 20(5):5538--5546, Feb 2012.

\bibitem{Cristiani:04}
Ilaria Cristiani, Riccardo Tediosi, Luca Tartara, and Vittorio Degiorgio.
\newblock Dispersive wave generation by solitons in microstructured optical
  fibers.
\newblock {\em Opt. Express}, 12(1):124--135, Jan 2004.

\bibitem{Dudley:2006nx}
John~M. Dudley, Go{\"e}ry Genty, and St{\'e}phane Coen.
\newblock Supercontinuum generation in photonic crystal fiber.
\newblock {\em Reviews of Modern Physics}, 78(4):1135--1184, 10 2006.

\bibitem{Efimov:03}
Anatoly Efimov, Antoinette Taylor, Fiorenzo Omenetto, Jonathan Knight, William
  Wadsworth, and Philip Russell.
\newblock Phase-matched third harmonic generation in microstructured fibers.
\newblock {\em Opt. Express}, 11(20):2567--2576, Oct 2003.

\bibitem{Belgiorno:10}
M.Clerici V.Gorini G.Ortenzi L. Rizzi E. Rubino V.G.Sala D.~Faccio F.Belgiorno,
  S.L.Cacciatori.
\newblock Hawking radiation from ultrashort laser pulse filaments.
\newblock {\em Phys Rev Lett}, 105(20), November 2010.

\bibitem{Fu2004}
Xiquan Fu, Liejia Qian, Shuangchun Wen, and Dianyuan Fan.
\newblock Nonlinear chirped pulse propagation and supercontinuum generation in
  microstructured optical fibre.
\newblock {\em Journal of Optics A: Pure and Applied Optics}, 6(11):1012--,
  2004.

\bibitem{Hawking1974}
S~Hawking.
\newblock Black-hole evaporation.
\newblock {\em Nature}, 248(5443):30--31, 1974.

\bibitem{Hawking1975}
S.W Hawking.
\newblock Particle creation by black holes.
\newblock {\em Communications in Mathematical Physics}, 43:199--220, 1975.

\bibitem{Herrmann:2002fj}
J.~Herrmann, U.~Griebner, N.~Zhavoronkov, A.~Husakou, D.~Nickel, J.~C. Knight,
  W.~J. Wadsworth, P.~St.~J. Russell, and G.~Korn.
\newblock Experimental evid2ence for supercontinuum generation by fission of
  higher-order solitons in photonic fibers.
\newblock {\em Physical Review Letters}, 88(17):173901--, 04 2002.

\bibitem{Husakou:2001rt}
A~V Husakou and J~Herrmann.
\newblock Supercontinuum generation of higher-order solitons by fission in
  photonic crystal fibers.
\newblock {\em Phys Rev Lett}, 87(20):203901, Nov 2001.

\bibitem{Birrell1984}
P.~C. W.~Davies N.~D.~Birrell.
\newblock {\em Quantum fields in curved space}.
\newblock Cambridge University Press, 1984.

\bibitem{Nation:2012uq}
P.~D. Nation, J.~R. Johansson, M.~P. Blencowe, and Franco Nori.
\newblock Colloquium: Stimulating uncertainty: Amplifying the quantum vacuum
  with superconducting circuits.
\newblock {\em Reviews of Modern Physics}, 84(1):1--24, 01 2012.

\bibitem{Paschotta}
R.~Paschotta.
\newblock {\em simulation software PROPULSE}.
\newblock RP Photonics Consulting GmbH, Zurich, Switzerland.

\bibitem{Philbin:2008fr}
Thomas~G. Philbin, Chris Kuklewicz, Scott Robertson, Stephen Hill, Friedrich
  K{\"o}nig, and Ulf Leonhardt.
\newblock Fiber-optical analog of the event horizon.
\newblock {\em Science}, 319(5868):1367--1370, 03 2008.

\bibitem{Price2003}
J.H.V. Price, T.M. Monro, K.~Furusawa, W.~Belardi, J.C. Baggett, S.~Coyle,
  C.~Netti, J.J. Baumberg, R.~Paschotta, and D.J. Richardson.
\newblock Uv generation in a pure-silica holey fiber.
\newblock 77(2-3):291--298--, 2003.

\bibitem{Robertson2011}
Scott~James Robertson.
\newblock {\em Hawking Radiation in Dispersive Media}.
\newblock PhD thesis, School of Physics and Astronomy University of St Andrews,
  2011.

\bibitem{Roy:2009kl}
Samudra Roy, S.~K. Bhadra, and Govind~P. Agrawal.
\newblock Dispersive waves emitted by solitons perturbed by third-order
  dispersion inside optical fibers.
\newblock {\em Physical Review A}, 79(2):023824--, 02 2009.

\bibitem{Rubino:2012fk}
E.~Rubino, A.~Lotti, F.~Belgiorno, S.~L. Cacciatori, A.~Couairon, U.~Leonhardt,
  and D.~Faccio.
\newblock Soliton-induced relativistic-scattering and amplification.
\newblock {\em Sci. Rep.}, 2, 12 2012.

\bibitem{Rubino:2012ly}
E.~Rubino, J.~McLenaghan, S.~C. Kehr, F.~Belgiorno, D.~Townsend, S.~Rohr, C.~E.
  Kuklewicz, U.~Leonhardt, F.~K{\"o}nig, and D.~Faccio.
\newblock Negative-frequency resonant radiation.
\newblock {\em Physical Review Letters}, 108(25):253901--, 06 2012.

\bibitem{Russell2003}
Philip Russell.
\newblock Photonic crystal fibers.
\newblock {\em Science}, 299(5605):358--362, January 2003.

\bibitem{Tianprateep2004}
M.~Tianprateep, Ji~Tada, T.~Yamazaki, and F.~Kannari.
\newblock Spectral-shape-controllable supercontinuum generation in
  microstructured fibers using adaptive pulse shaping technique.
\newblock {\em Jpn. J. Appl. Phys.}, 43:8059--8063, 2004.

\bibitem{Tianprateep2005}
Montian Tianprateep, Junji Tada, and Fumihiko Kannari.
\newblock Influence of polarization and pulse shape of femtosecond initial
  laser pulses on spectral broadening in microstructure fibers.
\newblock {\em Optical Review}, 12(3):179--189--, 2005.

\bibitem{Tran:2011qf}
Truong~X. Tran, Katiuscia~N. Cassemiro, Christoph S{\"o}ller, Keith~J. Blow,
  and Fabio Biancalana.
\newblock Hybrid squeezing of solitonic resonant radiation in photonic crystal
  fibers.
\newblock {\em Physical Review A}, 84(1):013824--, 07 2011.

\bibitem{Tu:12}
H.~Tu and S.A. Boppart.
\newblock Coherent fiber supercontinuum for biophotonics.
\newblock {\em Laser and Photonics Reviews}, 2012.

\bibitem{Tu:09}
Haohua Tu and Stephen~A. Boppart.
\newblock Optical frequency up-conversion by supercontinuum-free widely-tunable
  fiber-optic cherenkov radiation.
\newblock {\em Opt. Express}, 17(12):9858--9872, Jun 2009.

\bibitem{Tu:09b}
Haohua Tu and Stephen~A. Boppart.
\newblock Ultraviolet-visible non-supercontinuum ultrafast source enabled by
  switching single silicon strand-like photonic crystal fibers.
\newblock {\em Opt. Express}, 17(20):17983--17988, Sep 2009.

\bibitem{Unruh1981}
W.G. Unruh.
\newblock Experimental black-hole evaporation?
\newblock {\em Physical Review Letters}, 46:1351--1353, 1981.

\bibitem{Wai:86}
P.~K.~A. Wai, C.~R. Menyuk, Y.~C. Lee, and H.~H. Chen.
\newblock Nonlinear pulse propagation in the neighborhood of the
  zero-dispersion wavelength of monomode optical fibers.
\newblock {\em Opt. Lett.}, 11(7):464--466, Jul 1986.

\bibitem{Weinfurtner:2011uq}
Silke Weinfurtner, Edmund~W. Tedford, Matthew C.~J. Penrice, William~G. Unruh,
  and Gregory~A. Lawrence.
\newblock Measurement of stimulated hawking emission in an analogue system.
\newblock {\em Physical Review Letters}, 106(2):021302--, 01 2011.

\bibitem{Zhang2007}
Hua Zhang, Song Yu, Jie Zhang, and Wanyi Gu.
\newblock Effect of frequency chirp on supercontinuum generation in photonic
  crystal fibers with two zero-dispersion wavelengths.
\newblock {\em Opt. Express}, 15(3):1147--1154, 2007.

\bibitem{Zhu2004}
Zhaoming Zhu and Thomas Brown.
\newblock Effect of frequency chirping on supercontinuum generation in photonic
  crystal fibers.
\newblock {\em Opt. Express}, 12(4):689--694, 2004.

\end{thebibliography}

\end{document}